# Localized chemical switching of the charge state of nitrogen-vacancy luminescence centers in diamond

Toby W. Shanley, Aiden A. Martin, Igor Aharonovich[1, a)] and Milos Toth[1, b)]
School of Physics and Advanced Materials, University of Technology, Sydney, P.O. Box 123, Broadway, New South Wales 2007, Australia

We present a beam-directed chemical technique for controlling the charge states of near-surface luminescence centers in semiconductors. Specifically, we fluorinate the surface of H-terminated diamond by electron beam irradiation in the presence of $NF_3$ vapor. The fluorination treatment acts as a local chemical switch that alters the charge state of nitrogen-vacancy luminescence centers from the neutral to the negative state. The electron beam fluorination process is highly localized and can be used to control the emission spectrum of individual nanodiamonds and surface regions scanned by the electron beam.

Electron beams can be used to modify surfaces with high spatial resolution by irradiating solids in vacuum, or in low pressure, gaseous environments. Chemically active surface regions have previously been used as catalytic templates for electron beam directed, room temperature chemical vapor deposition,[1–3] and to locally accelerate the etch rate of diamond.[4] Here we present a beam-directed chemical technique for controlling the charge states of luminescence centers in semiconductors. Specifically, we fluorinate the surface of H-terminated diamond by electron beam irradiation in the presence of a precursor gas. The fluorination treatment acts as a local chemical switch that alters the charge state of near-surface nitrogen-vacancy (NV) luminescence centers from the neutral ($NV^0$) to the negative ($NV^-$) state.

The diamond NV center, comprised of a substitutional nitrogen atom adjacent to a carbon vacancy, is the subject of intense research into qubits for use in quantum computing,[5] markers for live cell imaging,[6,7] and sensors of extremely weak electric and magnetic fields.[8,9] These applications are made possible by the fact that the spin states of individual $NV^-$ centers can be manipulated deterministically and read out optically at room temperature,[10] and diamond is a non-toxic bio-compatible material.[11] However, the charge state of $NV^-$ centers can be unstable, and must be controlled to enable optimal exploitation in photonic and spintronic applications.[12,13] Shallow $NV^-$ centers located near a diamond surface are particularly significant for coupling NV centers to on-chip waveguides and cavities, and sensors that have high spatial resolution.[12]

The charge state of near-surface NV centers can be controlled by changing the electronegativity of molecules that terminate the diamond surface.[14–16] This is achieved here in a single step, with high spatial resolution by irradiating H-terminated diamond with an electron beam in the presence of $NF_3$ vapor. We show that, in the absence of electron irradiation, $NF_3$ does not alter H-terminated

———
a)Electronic mail: Igor.Aharonovich@uts.edu.au
b)Electronic mail: Milos.Toth@uts.edu.au

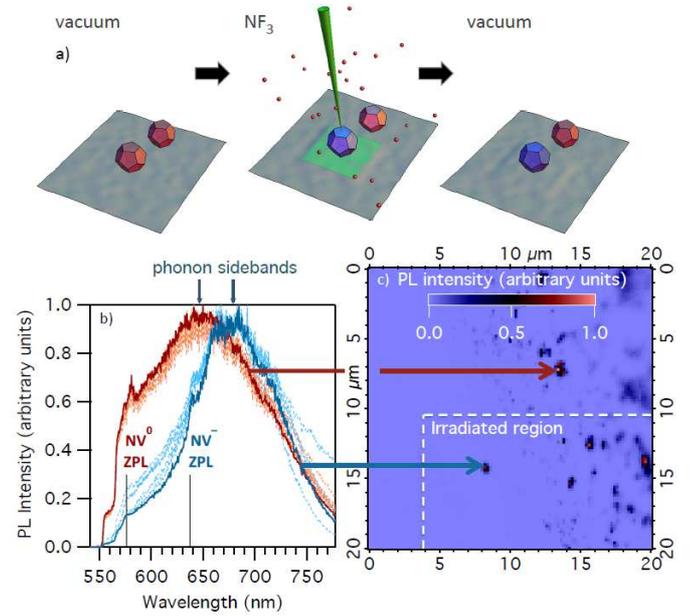

FIG. 1. (a) (Color online) Schematic illustration of nanodiamonds before, during and after electron irradiation under $NF_3$ vapor, b) PL spectra collected from irradiated and virgin nanodiamonds, and c) a PL map showing the sample region that was irradiated by an electron beam. Dashed red and blue traces show spectral characteristics of pre-irradiated and irradiated nanodiamonds respectively that have not been indicated on the map.

diamond provided it contains a protective overlayer of $H_2O$ molecules/fragments. Conversely, if the overlayer is desorbed (either thermally or through electron stimulated desorption), exposure to $NF_3$ at room temperature yields: (i) a stable change in photoluminescence (PL) caused by an increase in the ratio of negative to neutral NV centers; and, (ii) an increase in the ionization energy of diamond. The electron beam fluorination process is highly localized and can be used to switch the emission spectrum of individual nanodiamonds and surface regions scanned by the electron beam. The observed behavior is consistent with the effects of fluorine on band bending

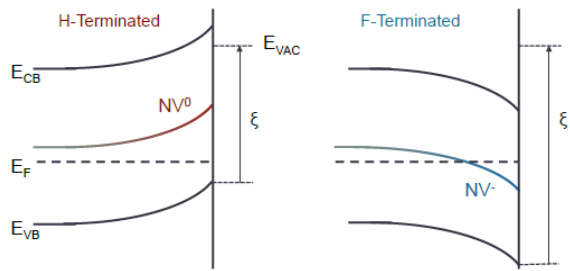

FIG. 2. Simplified schematic of a diamond surface terminated with hydrogen and fluorine. ($E_{CB}$ = conduction band minimum, $E_{VB}$ = valence band maximum, $E_F$ = Fermi level, $E_{VAC}$ = vacuum level, $\xi$ = ionization energy)

at the diamond surface, which are verified by measuring changes in the ionization energy of diamond using environmental photoemission yield spectroscopy (EPYS).

The electron beam process used to switch the charge state of NV centers is shown schematically in Fig. 1 (a). H-terminated nanodiamonds with a mean diameter of 30 nm were placed in a scanning electron microscope, and a selected region was irradiated by an electron beam in the presence of $NF_3$ vapor. The Irradiated and un-irradiated regions were examined subsequently by confocal PL.

The nanodiamonds were dispersed onto a Pt substrate and treated in an oxygen plasma (1 hour, 30 W, 30 Pa, 13.42 MHz) to remove hydrocarbon contaminants. Next, they were exposed to a hydrogen plasma (1 hour, 30 W, 30 Pa, 13.42 MHz) to produce hydrogen-terminated surfaces. The samples were then transferred in air to a scanning electron microscope with a base pressure of ~ 0.3 mPa. $NF_3$ vapor was delivered through a gas injection capillary that was positioned 150μm above the sample surface, and selected regions were irradiated with an electron beam at a background chamber pressure of 5 mPa (the local $NF_3$ pressure at the beam impact point on the sample surface was approximately three orders of magnitude greater, as explained elsewhere).[17] The electron beam energy and current were set to 20 keV and 0.51 nA respectively, and 20×20 μm² sample regions that could be identified in an ex-situ PL setup were irradiated for 60 minutes under these conditions. Panchromatic PL maps were collected in air at room temperature using a 532 nm excitation laser with a spot size of 0.25 μm², using a standard confocal microscope

PL spectra taken from unexposed and electron-irradiated regions of the sample are shown in Fig. 1 (b), along with a panchromatic PL map of the sample (c).The spectra show that, in the electron-irradiated region, the intensity of the $NV^-$ zero phonon line is greater than that of the corresponding $NV^0$ emission. Conversely, in the unexposed region, the intensity ratio is inverted. This difference is attributed to localized fluorination which causes downward band bending at the surface, and a change in the charge state of near-surface $NV^0$ centers, as shown schematically in Fig. 2.

The upward band bending shown in Fig. 2 is caused by an overlayer of $H_2O$ molecules and fragments on H-terminated diamond.[18] The slightly acidic water layer accepts electrons, producing a dipole field that bends the bands up at the surface. In contrast, the large electronegativity of fluorine inverts the polarity of the surface dipole, causing a downward bending of the bands in diamond.[19–21] The charge state of NV centers is determined by the position of the $NV^{0/-}$ charge transition level (shown in Fig. 2) relative to the Fermi level.[14] Near the surface, the transition level is located above and below the Fermi level in the case of hydrogen-terminated and fluorinated diamond, respectively. Fluorination therefore switches the charge state of near-surface $NV^0$ centers.

To confirm the above interpretation, we used EPYS to show that $NF_3$ can indeed give rise to the downward bending of energy bands shown in Fig. 2. Environmental photoemission yield spectroscopy was carried out using 5 × 5 mm², p-type polycrystalline diamond. A continuous diamond film was used for EPYS measurements (rather than nanodiamonds) to avoid spurious photoelectron emission from the substrate.

The polycrystalline diamond film was pre-treated with Oxygen and Hydrogen plasma under the same conditions as the nanodiamonds. EPYS Spectra were collected at a steady state pressure of $1.3 \times 10^2$ Pa by delivering $NF_3$ into the chamber at a flow rate of 1 sccm (the evacuation rate was controlled by pumping the chamber continuously through a needle valve). The detector bias was varied from +200 V to +400 V to optimize the cascade amplification in $H_2$ and $NF_3$ environments. The gas gain scales linearly with electron emission current, but is a function of gas type. The EPYS spectra shown are therefore normalized.

The EPYS technique enables the measurement of photoemission yield spectra in gaseous environments.[22] We used an EPYS system to: (i) obtain a photoemission spectrum at room temperature from H-terminated diamond in a $H_2$ environment; (ii) remove the $H_2O$ overlayer by evacuating the EPYS chamber to $6 \times 10^{-6}$ Pa and annealing the diamond in-situ at 100°C for one hour; and, (iii) obtain a photoemission spectrum at room temperature in an $NF_3$ environment. The resulting spectra (Fig. 3) show that H-terminated diamond has an electron emission threshold of 4.55 eV, and that it increased to 5.84 eV after annealing and exposure to $NF_3$ (the threshold remains at 4.55 eV if the annealing step is skipped and $H_2$ is simply replaced with $NF_3$. The threshold also does not change when the water overlayer is removed and $H_2$ is re-introduced, however in this case an increase in the photoelectron emission is observed). The ionization energy ($\xi$) of 4.55 eV is consistent with the negative electron affinity of H-terminated diamond.[23,24] The increase to 5.84 eV is consistent with the upward band bending shown in Fig. 2, but smaller than that of F-terminated (100) diamond, shown to be upwards of 8 eV.[20,21] The difference is attributed to incomplete fluorination, and the possibility that some surface sites are terminated by

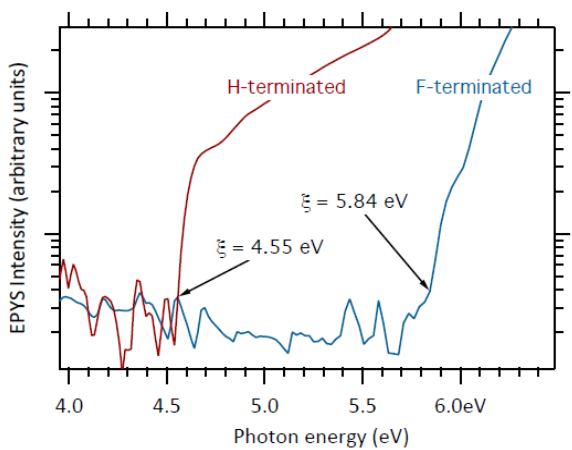

FIG. 3. (Colour online) Normalized photoemission yield spectra of H- and F-terminated diamond taken in $H_2$ and $NF_3$ environments. ($I_P$ = normalized photoelectron current, E = photon energy)

$NF_x$ species.

The fluorinated surface is stable upon exposure to air at room temperature. However, the photoemission threshold returns to 4.55 eV after heating to 100°C under vacuum, cooling to room temperature, and exposure to $H_2$. This behavior corresponds to thermal desorption of fluorine, which has been shown to occur over a wide temperature range, starting as low as 40°C.[25]

The changes in ionization energy observed in the EPYS experiments are consistent with the band bending schematic shown in Fig. 2. A downward shift of the valance band maximum relative to the vacuum level is expected to cause a corresponding increase in electron emission threshold, and to switch the charge state of near-surface $NV^0$ centers. We attribute the localized, electron-beam-induced version of the process to electron beam stimulated desorption of $H_2O$ molecules and fragments from H-terminated diamond, and subsequent exchange of hydrogen with fluorine. The fluorination step can take place spontaneously at room temperature (as shown by EPYS), but is likely assisted by electron-beam-induced decomposition[17] of $NF_3$ adsorbates.

We note that prior methods for fluorinating diamond have included exposure to elemental and molecular fluorine,[26] $XeF_2$,[21] $CF_4$ plasma,[27] and $CHF_3$ plasma.[28] However, all of these methods are delocalized, and most are prone to the generation of defects that compromise the optoelectronic properties of diamond. The room temperature, $NF_3$-mediated, electron beam technique described here is minimally invasive and enables highly localized, direct-write processing of diamond.

In conclusion, fluorination of H-terminated diamond is achieved by exposure to $NF_3$ after the desorption of a water layer. Desorption was achieved either by thermal annealing to 100°C in ultra-high vacuum, or by electron beam irradiation of diamond in a gaseous $NF_3$ environment. Fluorination causes a change in the PL spectrum characteristic of an increase in the ratio of negative to neutral NV centers, and an increase in the ionization energy of diamond. Both results are consistent with an energy band model of hydrogen-terminated and fluorinated diamond.

A.A.M. is the recipient of a John Stocker Postgraduate Scholarship from the Science and Industry Endowment Fund. This work was supported by FEI Company and the Australian Research Council (Project Number DP140102721). I.A. is the recipient of an Australian Research Council Discovery Early Career Research Award (Project Number DE130100592). We thank Philippe Bergonzo and Huan-Cheng Chang for the diamond samples used in this work.